\documentclass[preprint,showpacs,preprintnumbers,amsmath,amssymb]{revtex4}

\usepackage{graphicx}
\usepackage{dcolumn}
\usepackage{bm}

\begin{document}

\preprint{APS/123-QED}

\title{ Assesment of multifragmentation under the effect of symmetry energy and cross-section\\}

\author{Rubina Bansal}
\author{Suneel Kumar}%
 \email{suneel.kumar@thapar.edu}
\affiliation{%
School of Physics and Material Science, Thapar University, Patiala-147004, Punjab (India)\\
}%
\date{\today}
\begin{abstract}
The effect of symmetry energy and cross section had seen on the fragment production in the multifragmentation of $^{20}Ne_{10}+ ^{20}Ne_{10}$ and $^{197}Au_{79} + ^{197}Au_{79}$ at incident energy 50-1000 MeV/nucleon using isospin dependent quantum molecular dynamics model. To see the effect of symmetry energy and cross-section in more effective way relative yield of $^{197}Au_{79}$ and $^{20}Ne_{10}$ was studied. It is observed that relative yield of free nucleon, light mass fragment at isospin dependent cross-section is almost same at different symmetry energy values. While for fixed cross-section relative yield is influenced by symmetry energy. We compare the stiffness of symmetry energy suggested by different groups with our value. To verify our result we compare our study with ALADIN data. 
\end{abstract}
\maketitle
\baselineskip=1.5\baselineskip\
\section{Introduction}
 The science of nuclear physics deals with the properties of nuclear matter. The study of nature of matter and strength of nuclear interaction is key to understand many fundamental problems. Understanding the behaviour of nuclear matter at density and isospin away from normal nuclear matter (T $\approx$ 0 MeV, $\rho_{0}$ $\approx$ 0.16 $fm^{-3}$, N $\approx$ Z ) has gained tremendous importance \cite{isospin}. In the past decade only light ions or particles could be accelerated to produce the collision. But it is possible to accelerate the heavy ions to a large amount of energy. The term heavy ion is generally used for nuclei which are heavier than the helium nucleus. The heavy ion collisions has attracted the scientific world because of its various features. The intermediate energy heavy ion collisions are helpful in studying the nuclear matter at extreme conditions of temperature and pressure, which can be correlated with astrophysical happening like supernova explosion, neutron star etc.\\ 
                            Multifragmentation is one of the important phenomena occurring at intermediate energies. The process of breaking of colliding nuclei into several small, medium size fragments is called multifragmentation. Those fragments which are highly excited and neutron-rich subsequently undergo de-excitation to cold and stable isotopes. Similar hot nuclei are also produced in the interior of a collapsing star and subsequent supernova explosion \cite{mishustin}.\\
 The production of these nuclei depends on their internal excitation and is sensitive to the symmetry energy part of the binding energy \cite{lozhkin}. Puri and co-workers \cite{vinayak} has analysed the time evolution of fragment production for symmetric nuclei Ca + Ca, Xe + Xe and Au + Au. There are many parameters which effects the multifragmentation. Symmetry energy is one of the most promising parameter to study the multifragmentation after the collision between two nuclei. Symmetry energy $E_{sym}(\rho)$ of nuclear matter characterizes how the energy changes as one move away from equal numbers of neutrons and protons. In the heavy ion collisions the dynamics of the collisions between any two nuclei is also sensitive to isospin dependent NN cross section. So, it is an interesting and important goal of heavy ion physics to extract information of symmetry energy and its density dependence.  Many groups had found the relation of symmetry energy \cite{Zhao, chen, Shetty}. They consider different values of $E^{0}_{sym}$ and $\gamma$ and see the influence of stiffness and softness of the symmetry energy. The best estimate of the density dependence of the symmetry energy can be parameterized as ref. \cite{dd}
\begin{equation}
 E_{sym}= E^{0}_{sym}(\frac{\rho}{\rho_{0}})^\gamma
\end{equation}
In multifragmentation the measurement of fragment isotopic yield distribution can provide important insight into the symmetry energy and the decay characteristics of these nuclei. It has been shown in experimental measurements that the ratio of the fragment yields, $R_{12}$ (N, Z), taken from two different multifragmentation reactions, 1 and 2, obeys an exponential dependence on the neutron number (N) and the proton number (Z) of the fragments; an observation known as isoscaling \cite{Tsang}. The dependence is characterized by the relation
\begin{equation}
R_{21}(N,Z) = Y_{2}(N,Z)/Y_{1}(N,Z) = Ce^{({\alpha}N + {\beta}Z)}                                   
\end{equation}                               
Where $Y_{2}$ and $Y_{1}$ are the fragment yields from the neutron-rich and the neutron-deficient systems, respectively, C is an overall normalization factor, and $\alpha$ and $\beta$ are the parameters characterizing the isoscaling behaviour. So far, the isoscaling behaviour has been studied experimentally and theoretically for different reaction mechanisms \cite{Ma, Fang, Tian}. D. V. Shetty {\it et al.,} \cite{shetty2, shetty3} , experimentally understand the correlation between the temperature, density and symmetry energy of multifragmentating system as it evolves with excitation energy. They also study the relative reduced neutron and proton densities as a function of excited energy of fragmenting source for the $^{58}Fe$ + $^{58}Ni$ and $^{58}Fe$ + $^{58}Fe$ reaction. Their study shows a steady decrease in reduced neutron density and an increase in proton density with increase excitation energy. Two nucleons can collide if they come closer than certain distance. Multifragmentation is also influenced by cross section. The nuclear cross-section is used to characterize the probability that a NN collision will occur. C. Zeitlin {\it et al.,} \cite{Zeitlin} experimentally study the effect of cross section on fragment production. R. Wada and co-worker \cite{Wada} shows that value of cross-section decrease with increase in incident energy.\\ 
In this paper we compare the various forms of symmetry energy suggested by different groups with our suggested value and see its softness and stiffness for different values of the parameter $\gamma$. We have tried to understand the influence of symmetry energy on multiplicity of fragments for neutron rich and deficient nuclei at different incident energies i.e. we see the effect of mass dependence on fragment production under the effect of symmetry energy. In addition to symmetry energy we analyse the relative production of fragment FN (A = 1) and LMF (2 $\le$ A $\le$ 4)) under the effect of cross section ( isospin dependent and fixed (20 and 50 mb)cross section). To verify the result we compare our calculations with experimental ALADIN data \cite{Schtittauf, tsang2}. This study is done within the frame work of isospin quantum molecular dynamics model \cite{10}. Section II deals with the model, Sec. III discusses the results.\\
\section{ISOSPIN-DEPENDENT QUANTUM MOLECULAR DYNAMICS (IQMD) MODEL}
Our study is performed within the framework of IQMD \cite{10} model, which is an improved version of the QMD model \cite{11} developed by J. Aichelin and coworkers and then improved by Puri and coworkers has been applied to explain various phenomenon such as collective flow, disappearance of flow, fragmentation and elliptical flow successfully. The isospin degree of freedom enters into the calculations via symmetry potential, cross-sections and Coulomb interactions \cite{10}. The details about the elastic and inelastic cross-sections for proton-proton and neutron-neutron collisions can be found in ref. \cite{10}. In IQMD model, the nucleons of target and projectile interact via two and three-body Skyrme forces, Yukawa potential and Coulomb interactions. In addition to the use of explicit charge states of all baryons and mesons, a symmetry potential between protons and neutrons corresponding to the Bethe- Weizsacker mass formula has been included. Skyrme forces are very successful in the analysis of low energy phenomena such as fusion, fission and cluster-radioactivity, where nuclear potential plays an important role \cite{24}. \\
where hadrons propagate with Hamilton equations of motion:
\begin{equation}
\frac{d{r_i}}{dt}~=~\frac{d\it{\langle~H~\rangle}}{d{p_i}}~~;~~\frac{d{p_i}}{dt}~=~-\frac{d\it{\langle~H~\rangle}}{d{r_i}},
\end{equation}
with
\begin{eqnarray}
\langle~H~\rangle&=&\langle~T~\rangle+\langle~V~\rangle\nonumber\\
&=&\sum_{i}\frac{p_i^2}{2m_i}+
\sum_i \sum_{j > i}\int f_{i}(\vec{r},\vec{p},t)V^{\it ij}({\vec{r}^\prime,\vec{r}})\nonumber\\
& &\times f_j(\vec{r}^\prime,\vec{p}^\prime,t)d\vec{r}d\vec{r}^\prime d\vec{p}d\vec{p}^\prime .
\end{eqnarray}
 The baryon-baryon potential $V^{ij}$, in the above relation, reads as:
\begin{eqnarray}
V^{ij}(\vec{r}^\prime -\vec{r})&=&V^{ij}_{Skyrme}+V^{ij}_{Yukawa}+V^{ij}_{Coul}+V^{ij}_{sym}\nonumber\\
&=& \left [t_{1}\delta(\vec{r}^\prime -\vec{r})+t_{2}\delta(\vec{r}^\prime -\vec{r})\rho^{\gamma-1}
\left(\frac{\vec{r}^\prime +\vec{r}}{2}\right) \right]\nonumber\\
& & +~t_{3}\frac{exp(|\vec{r}^\prime-\vec{r}|/\mu)}{(|\vec{r}^\prime-\vec{r}|/\mu)}~+~\frac{Z_{i}Z_{j}e^{2}}{|\vec{r}^\prime -\vec{r}|}\nonumber\\
& & + t_{6}\frac{1}{\varrho_0}T_3^{i}T_3^{j}\delta(\vec{r_i}^\prime -\vec{r_j}).
\label{s1}
\end{eqnarray}
Here $Z_i$ and $Z_j$ denote the charges of $i^{th}$ and $j^{th}$ baryon, and $T_3^i$, $T_3^j$ are their respective $T_3$
components (i.e. 1/2 for protons and -1/2 for neutrons). Meson potential consists of Coulomb interaction only.
The binary nucleon-nucleon collisions are included by employing the collision term of well known VUU-BUU equation.
 During the propagation, two nucleons are
supposed to suffer a binary collision if the distance between their centroids
\begin{equation}
|r_i-r_j| \le \sqrt{\frac{\sigma_{tot}}{\pi}}, \sigma_{tot} = \sigma(\sqrt{s}, type),
\end{equation}
"type" denotes the ingoing collision partners (N-N, N-$\Delta$, N-$\pi$,..). In addition,
Pauli blocking (of the final
state) of baryons is taken into account by checking the phase space densities in the final states. The phase space generated using IQMD model has been analysed using minimum spanning tree algorithm \cite{24}\\
\section{Results and Discussion}
We have simulated 1000 events involving the symmetric reactions of $^{20}Ne_{10}+ ^{20}Ne_{10}$ and $^{197}Au_{79} + ^{197}Au_{79}$ at incident energies 50, 100, 200, 400, 600, 1000 MeV/nucleon for central collision (scaled impact parameter is 0.3) using soft equation of state. Here, we use three different forms of symmetry energies. For all the calculations we take symmetry energy corresponding to normal density is 32 MeV and value of $\gamma$ which characterizes the stiffness of the symmetry energy is 0, 0.66 and 2. In figure 1, we compare the stiffness of symmetry energy suggested by different groups.\\
\begin{figure}
\hspace{-2.0cm}\includegraphics[scale=0.45]{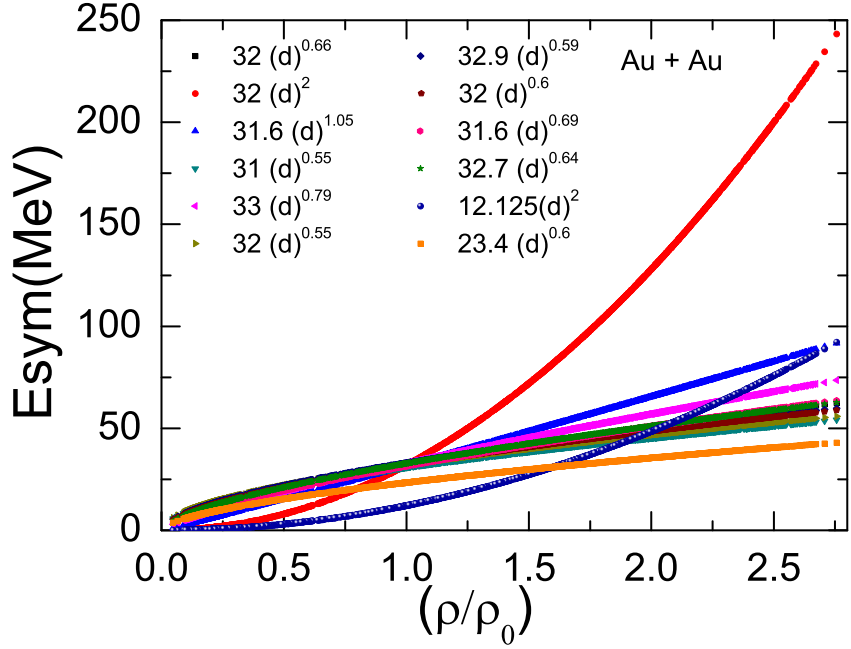}\caption{\label{Figure:1}Schematic diagram to show the stiffness of symmetry energy suggested by various groups.}
\end{figure}
Red and royal coloured symbol having $\gamma$ = 2 are highly stiffed than others. Here we analyse that larger the $\gamma$ value more the stiffness. For figure 1 density dependence of symmetry energy obtained from various groups shows the close agreement with parameterized forms of the density dependence of the symmetry energy, given as in eq. (1), where $E^{0}_{sym}$ $\approx$ 31-32 MeV and $\gamma$ $\approx$ 0.55-0.69 which is agree with ref. \cite{shetty2}.   
\begin{figure}
\hspace{-2.0cm}\includegraphics[scale=0.45]{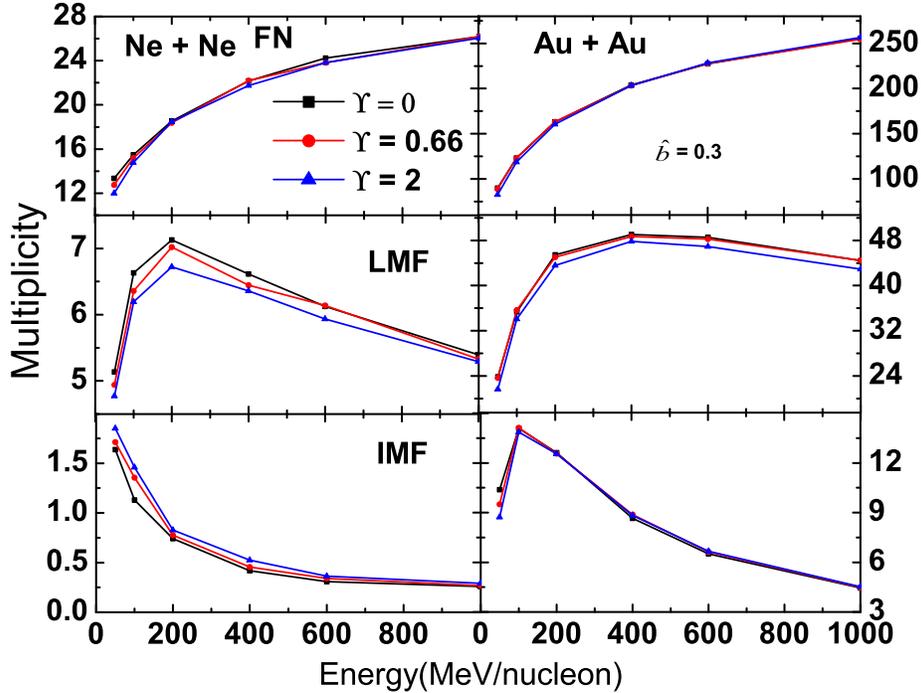}
\caption{\label{Figure:2} Influence of symmetry energy on fragment production for symmetric reaction Au + Au and Ne + Ne at different energies for central collision ( $\hat{b}$=0.3).}
\end{figure}
We study the influence of symmetry energy on fragment production at different incident energies in figure 2. On the basis of mass number the fragments are classified as free nucleons [ A = 1 ], light mass fragments (LMFs) [2 $\le$ A $\le$ 4], intermediate mass fragments (IMFs) [5 $\le$ A $\le$ $A_{tot}$ / (3 or 6)].\\ 
      From the Figure 2, we observed that the multiplicity of FN $\rangle$ LMF $\rangle$ IMF for both neutron rich and deficient nuclei. On considering the mass effect for lightest nuclei (Ne + Ne) at low energy (50 MeV/nucleon) the relative ratio of FN: LMF: IMF is 8: 3: 1 and for high energy (1000 MeV/nucleon) is FN: LMF: IMF :: 7: 2: 1. Due to less Coulomb effect in the collision of lighter nuclei the effect of incident energy is almost same as one move from low energy to high energy. While in case of heaviest nuclei (Au+Au) relative ratio of fragment production at low energy is FN: LMF: IMF :: 9: 3: 1 and at high energy FN: LMF: IMF :: 83: 9: 1. There is large change in the formation of fragment, due to large compression at high energy and Coulomb repulsion. We study that the rate of change of fragment production is decrease as we go beyond 400 MeV/nucleon. It is also experimentally observed that a little change in fragmentation yield takes place beyond 400 MeV/nucleon \cite{Schtittauf}. The slope of fragment production in all the cases below below saturation time is high due to the Pauli blocking.\\
 On considering the influence of symmetry energy on fragment production, we analyse FN for neutron rich nuclei is almost independent from the affect of symmetry energy. But one can see a measureable effect of symmetry energy for neutron deficient nuclei at 100 MeV/nucleon. In LMF production influence of symmetry energy is large as compared to FN. We got pronounced effect of different symmetry energies in both the cases as shown in second horizontal panel. On observing IMF's production for light nuclei we can say IMF multiplicity increases with increase in the stiffness of symmetry energy. But for heavy nuclei production of IMF is independent of symmetry energy and we observed measureable effect of symmetry energy only at low energy.  While for light nuclei we see the influence of symmetry energy at all energy range as shown in figure.\\
                             So far we have seen the effect of incident energy, symmetry energy on fragment production. To see the effect of symmetry in more effective way we include the isoscaling parameter and cross-section. Here we fix cross-section at 20 and 55 mb to study the effect of multifragmentation. For comparison, we shall also use an isospin dependent cross-section where $\sigma_{np}$ is more as compare to $\sigma_{pp}$, $\sigma_{nn}$. We see the effect of cross-section on production of FN and LMF for two different multifragmentation reactions i.e. ratio of production of FN and LMF studied between a neutron rich nuclei Au and a neutron deficient nuclei Ne for incident energies 50, 100, 200, 400, 600, 1000 MeV/nucleon for different symmetry energy.\\ 
                                       In first panel we consider isospin dependent cross-section. In second and third panel we use constant isospin independent cross-section with value 20 mb and 55 mb respectively. One can see that fragment production in addition to symmetry energy is affected by different cross-section. \\
 \begin{figure}
\hspace{-2.0cm}\includegraphics[scale=0.45]{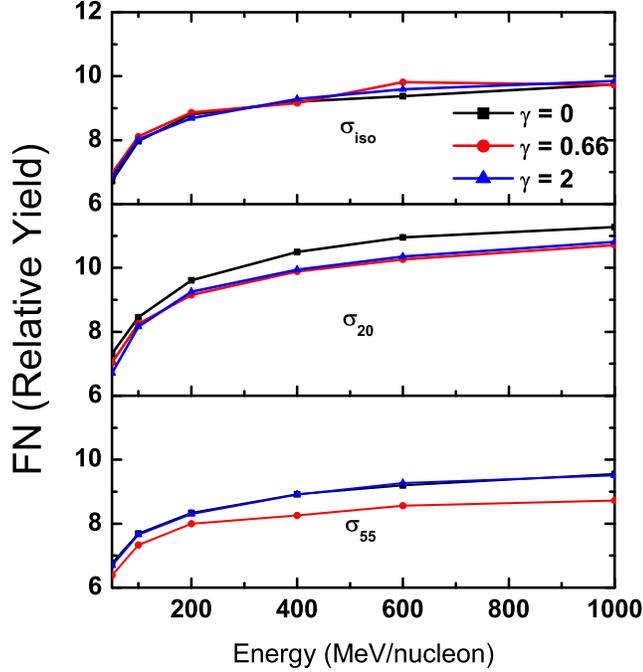}
\caption{\label{Figure:3} Effect of cross-section with incident energy on production of free nucleon at different symmetry energy between a neutron rich nuclei Au and neutron deficient nuclei Ne.}
\end{figure}
In the figure 3 and 4 we display the relative production of FN is more than LMF's in all the cases of isospin dependent and fixed cross-section. We can justify our study on the basis of the Cugnon parameterization \cite{huang}. For the energy range 50 to 1000 MeV/nucleon, elastic collisions take place. In figure 2 we see the influence of symmetry energy on production of FN and LMF, but when we include the cross section, the slope of relative ratio of neutron rich and deficient nuclei decrease with increase in incident energy as shown in fig. 3 and 4. Moreover the role of cross-section decreases with increase in energy. The behaviour of relative yield curve at $\gamma$ = 2 shift due to the energy effect. the relative yield ratio at $\sigma_{iso}$ and $\sigma_{55}$ is approximately 10 at high energy, while at $\sigma_{20}$ its value decrese for FN. For LMF the relative ratio at $\sigma_{iso}$ and $\sigma_{55}$ is approximately 5 at low energy and 8 for high energy.\\  
\begin{figure}
\hspace{-2.0cm}\includegraphics[scale=0.45]{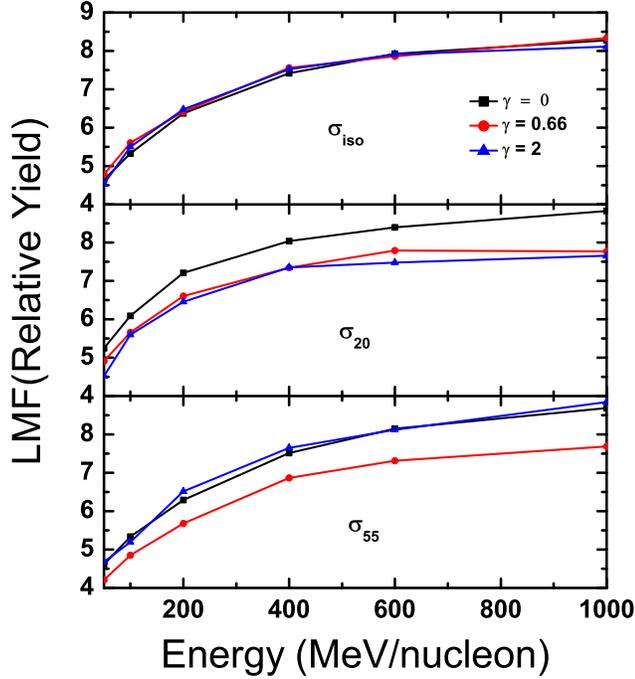}
\caption{\label{Figure:4} Effect of cross-section with incident energy on production of light mass fragments at different symmetry energy between a neutron rich nuclei Au and neutron deficient nuclei Ne.}    
\end{figure}
In figure 5 we compare our study of IMF's production at different energies of Au + Au collision with experimental data of ALADIN for central collisions \cite{Schtittauf, tsang2}. The production of IMF's decrease with incident energy, because at high energy the nuclear matter shatter down in small pieces which leads to more production of FN and LMF at the cost of decrease in the production of IMF's. In figure 5 we have displayed the results at cross-section 20 mb, 55 mb, and $\sigma_{iso}$ in the absence ($\gamma$ = 0) and presence ($\gamma$ = 0.66) of density dependence of symmetry energy. One can see that although the theoretical results are in not in close agreement with the data, but the trend of result in similar to experimental finding. The mismatch in the trend may be due to the binding energy and momentum parameter. As stated earlier cluster and fragments in this study were generated by simple minimum spanning tree method. This method binds two nucleons based on spatial correlation regardless of the fact whether proper binding energy is achieved or not. The new methods based on binding energy creteria can shed light on this aspect \cite{x}. It was noted by many authors \cite{y} that the various forms of equation of state as well as momentum dependent interactions can also make different impact on the results.  \\
\begin{figure}
\hspace{-2.0cm}\includegraphics[scale=0.45]{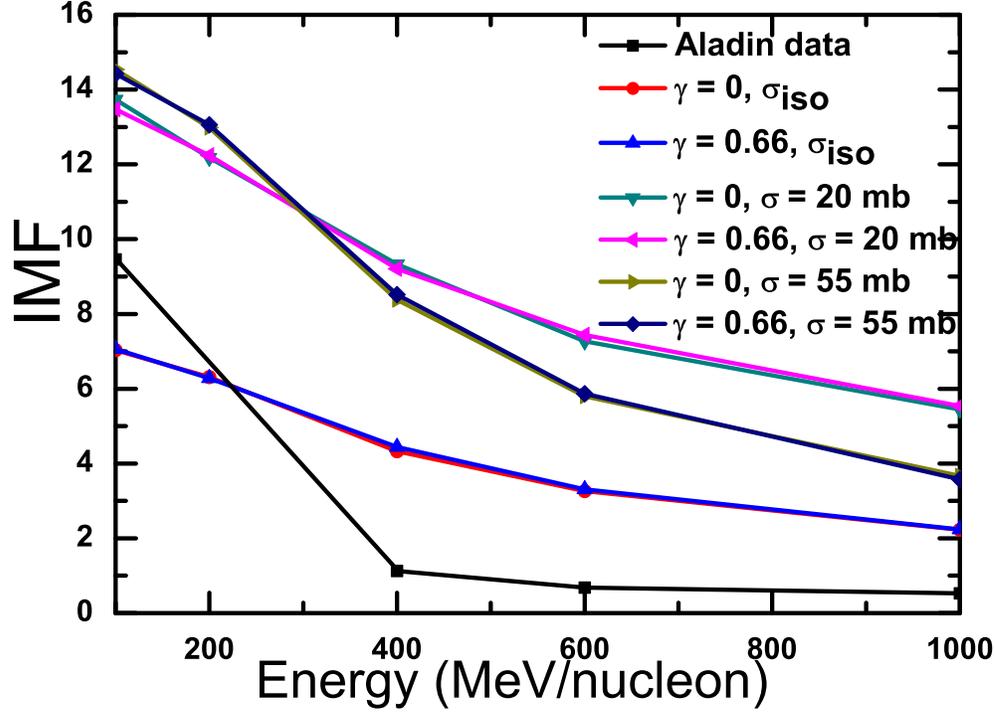}
\caption{\label{Figure:5} Comparison of average multiplicity of intermediate mass fragments (IMF's) with ALADIN data  at incident energies of 100, 200, 400, 600, 1000 MeV/nucleon as a function of energy.}    
\end{figure}
\section{Summary}
 In this paper we have used isospin dependent quantum molecular dynamics model to study the influence of symmetry energy and cross-section on multifragmentation at neutron rich (Au) nuclei as well as neutron deficient (Ne) nuclei at incident energy 50-1000 MeV/nucleon. All the simulations carried at central mass frame with scaled impact parameter 0.3. We used momentum independent and soft equation of state. Our study showed that the  multiplicity of fragments depends upon incident energy, symmetry energy influences the fragment production, effect of symmetry energy is more for neutron deficient nuclei than the neutron rich nuclei and cross section affects the fragment production.\\ 


\begin{thebibliography}{999}
\bibitem{isospin} Isospin Physics in Heavy Ion Collision at intermediate Energies, edited by B. A. Li and W. Schroder ( Nova Science, New York, 2011); B.A. Li, L.W. Chen, C. M. Ko, Phys. Rep. {\bf 464}, 113(2008).
\bibitem{mishustin} A. S. Botvina and I. N. Mishustin, Phys. Lett. {\bf B 584}, 233(2004).
\bibitem{lozhkin} A. S. Botvina, O. V. Lozhkin, and W. Trautmann, Phys. Rev. {\bf C 65}, 044610(2002).
\bibitem{vinayak} K. S. Vinayak and S. Kumar, Phys. Rev. {\bf C 83}, 034614(2011). 
\bibitem{Zhao} Zhao- Qing Feng, Phy. Rev. {\bf C 84}, 024610(2011).
\bibitem{chen} L. W. Chen, C. M. Ko. And B. A. Li, Phys. Rev. Lett. {\bf 94}, 032701(2005).
\bibitem{Shetty} D. V. Shetty, S. J. Yennello and G. A. Souliotis, Phys. Rev. {\bf C 75}, 034602(2005).
\bibitem{dd} L. W. Chen, C. M. Ko, B. A. Li, Phys. Rev. Lett {\bf 94}, 032701(2005).
\bibitem{Tsang} M. B. Tsang, W. A. Friedman, C. K. Gelbke, W. G. Lynch, G. Verde, and H. S. Xu, Phys. Rev. Lett {\bf 86}, 5023(2011). 
\bibitem{Ma} Ma Yu-Gang et al. Phys. Rev. {\bf 69}, 065610(2005).
\bibitem{Fang} Fang De-Qing {\it et al.,} J. Phys. {\bf 34}, 2173(2007).
\bibitem{Tian} Tian Went-Dong {\it et al.,} Phys. Rev. {\bf 76}, 024607(2007). 
\bibitem{shetty2} D. V. Shetty, S. J. Yennello, and G. A. Souliotis Phys. Rev. {\bf C 76}, 024606(2007).
\bibitem{shetty3} D. V. Shetty, A. S. Botvina, S. J. Yennello, G. A. Souliotis, E. Bell, A. Keksis Phys Rev. {\bf C 71}, 024602(2005).
\bibitem{Zeitlin} C. Zeitlin {\it et. al.,} Phys. Rev. {\bf C 56}, 388(1997).
\bibitem{Wada} R. Wada {\it et al.,} Phys. Rev. {\bf C 69}, 044610(2004).
\bibitem{Schtittauf} A. Schtittauf {\it et al.,} Nucl. Phys. {\bf A 607}, 457(1996).
\bibitem{tsang2} M.B. Tsang {\it et al.,} Phys. Rev. Lett. {\bf 71}, 1502(1993).
\bibitem{10} C. Hartnack {\it et al.,} Eur. Phys. J. {\bf A 1}, 151(1998); S. Gautam {\it et al.,} J. Phys. {\bf G 37}, 085102(2010): S. Gautam {\it et al.,} Phys. Rev. {\bf C 83}, 014603(2011);{\it ibid} {\bf C 83},034606(2011). {\it ibid} Phys. Rev. {\bf C 83} 034606(2011);  V. Kaur and S. Kumar Phys. Rev. {\bf C 81}, 064610(2010); {\it ibid } Nucl. Phys. {\bf A 861}, 37(2011); V. Kaur, S. Kumar and R. K. Puri Phys. Lett. {\bf B 697}, 512(2011), S. Kumar, Rajni and S. Kumar Phys. Rev. {\bf C 82}, 024610(2011).   
\bibitem{11} J. Aichelin Phys. Rep. {\bf 202}, 233(1991); E. Lehmann Phys. Rev. {\bf C 51}, 2113(1995); {\it ibid} Prog. Nucl. Part. Phys. {\bf 30}, 219(1993), S. Kumar {\it et al.'} Phys. Rev. {\bf C 57}, 2744(1998); S. Goyal {\it et. al.,} Nucl. Phys. {\bf A 853}, 164(2011); {\it ibid} {\bf C 83}, 047601(2011). S. Kumar and R. K. Puri Phys. Rev. {\bf C 58}, 320(2011).
\bibitem{24} R. K. Puri {\it et al.,} Phys. Rev. {\bf C 45}, 1837(1997); ibid {\bf 43}, 315(1991); {\it ibid} Eur. Phys. J. {\bf A 23}, 429(2005); {\it ibid} {\bf A 3}, 277(1998); {\it ibid} {\bf A 8}, 103(2000), I. Dutt {\it et al.,} Phys.
Rev. {\bf C 81}, 044615(2010); {\it ibid}, 047601(2010); {\bf ibid}, 064609(2010); {\bf ibid}, 064608(2010); S. Kumar {\it et al.,} Chineese Phys. Lett. {\bf 27}, 062504(2010).  
\bibitem{huang} S. W. Huang, Ph.D thesis, Tubingen (Germany) (1994).
\bibitem{x} R. K. Puri, C. Hartnack and J. Aichelin, Phys. Rev. {\bf C 54}, 28(1996); R. K. Puri and J.Aichelin, J. Comp. Phys. {\bf 162}, 245(2000); Y. K. Vermani and R. K. Puri, Eur. Phys. Lett. {\bf 85}, 062001(2009); ibid J. Phys. {\bf G 36}, 0105103(2009); ibid {\bf 37}, 015105(2010).
\bibitem{y} Y. K. Vermani, S. Goyal and R. K. Puri, Phys. Rev. {\bf C 79}, 064613(2009); A. D. Sood and R. K. Puri, Phys. Rev. {\bf C 79}, 064618(2009); ibid Eur. Phys. J. {\bf A 30}, 571(2006); R. K. Puri {\it et al}., Nucl. Phys. {\bf A 575}, 733(1994); R. Chugh and R. K. Puri, Phys. Rev. {\bf C 82}, 014603(2010); S. Kumar, S. Kumar and R. K. Puri, Phys. Rev. {\bf C 78}, 064602(2008).  
\end{thebibliography}
\end{document}